# Statistical analysis of many single-molecule encounters reveals plasmonic resonance dependent nanoantenna-molecule interactions

Lisa Saemisch[1], Matz Liebel[1] and Niek van Hulst[1,2]

[1] ICFO-Institut de Ciencies Fotoniques, The Barcelona Institute of Science and Technology, 08860 Castelldefels, Barcelona, Spain

[2] ICREA-Institució Catalana de Recerca i Estudis Avançats, 08010 Barcelona, Spain

The nanoscale interaction between single emitters and plasmonic structures is traditionally studied by relying on near-perfect, deterministic, nanoscale-control. This approach is ultra-low throughput thus rendering systematic studies difficult to impossible. Here, we show that super resolution microscopy in combination with data-driven statistical analysis allows studying near-field interactions of single molecules with resonant nanoantennas. We systematically tune the antennas' spectral resonances and show that emitters can be separated according to their coupling strength with said structures which ultimately allows the reconstruction of 2D interaction maps around individual nanoantennas.

The advent of super-resolution techniques has revolutionised fluorescence microscopy over the past decades[1–4]. Especially the field of biology has dramatically benefited from the possibility to study structures with resolution far below the diffraction limit. Here, super-resolution microscopy (SRM) allows studying single molecules[5,6], supramolecular assemblies[7] and, with ongoing experimental optimisation, even whole cells and small organisms[8]. Given its success in biology and ease of implementation it comes as no surprise that other fields are starting to exploit its unique capabilitites.

Cang *et al.*[9] were the first to use SRM to map sub-diffraction limited features on plasmonic substrates. They interrogated a rough gold film using the Brownian motion of freely diffusing molecules in solution and resolved surface-regions of strong fluorescence enhancement which they attributed to hotspots of approximately 15 nm in size. Their experimental interpretations were based on the assumption that determining the precise location of a fluorescence emission event is equivalent to determining an emitter's nanometric position. However, this simple assumption is based on isotropically emitting dipoles and, depending on the experimental parameters, might be only partially valid[10]. Given that far-field localizations are strongly dependent on the mere orientation of an emitter it comes as no surprise that placing a resonant nanostructure next to a molecule can dramatically alter its apparent position. This phenomenon has received considerable attention and the apparent position of molecules close to different metallic nanostructures, such as triangles[11], nanorods[12,13], nanowires[14] and more complex trimer structures[15] has been studied extensively with SRM. Unsurprisingly, all of these works show discrepancies between the true single-emitter position and its super-localized position, determined in the far field, as expected for weak emitters placed in proximity to a resonant plasmonic structure with a strong dipole-moment[16,17].

Despite these imaging problems, the combination of plasmonics with SRM holds great promise for systematically studying the interaction between individual nanostructures and single emitters. Traditionally, near-field scanning optical microscopy (NSOM) has been used to study these interactions[18–21]. Here, tailor-made nanostructures probe single-emission sites, such as single molecules immobilised on a glass surface, and the near-field coupling between the two entities is obtained by deterministically varrying the structure's position (Figure 1a). Albeit the ultimate control offered by NSOM, the experimental complexity of this approach dramatically limits its applicability and throughput. As an alternative to NSOM one can imagine using a solution of freely diffusing molecules which stochastically sample the immobilised nanostructures (Figure 1b). Here, near-field phenomena can be explored as long as sufficient single-molecule events, alongside their positions, are obtained. Even though offering less control than the determinstic single-point NSOM, the multiplexing capabilities of plasmonic SRM are dramatic thus, in principle, allowing the simultaneous study of hundreds of individual nanostructures (Figure 1c). Experimentally, one simply mounts the nanoantennas of interest on top of a widefield microscope, submerges the structures in a solution containing the fluorescent probe-molecules of choice and records thousands of fluorescence encounters over the course of a few hours (Figure 1d,e).

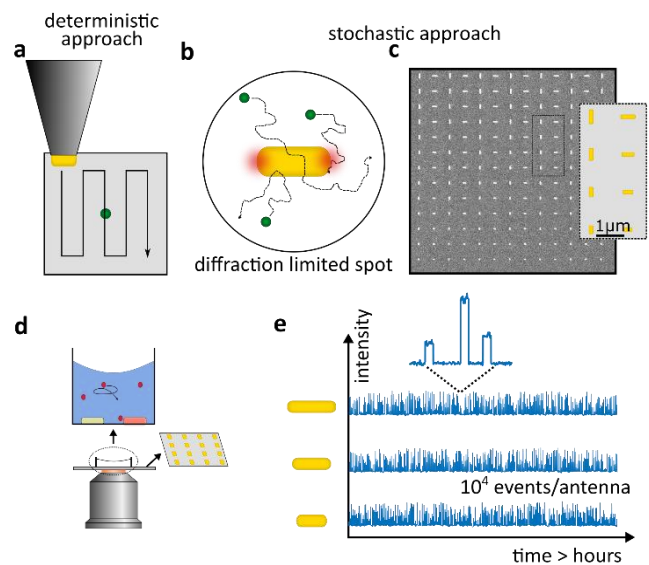

**Figure 1, Principle of a stochastic nanoscale interaction experiment**. a) In a typical NSOM experiment a nanorod is deterministically scanned over a single emitter in order to study the mutual interaction. b) Freely diffusing molecules stochastically scan a metallic nanostructure with sub-diffraction limited resolution. c) Typical sample geometry comprised of a large array of indvidual nanoantennas (SEM image alongside schematic sample lay-out). d) Cartoon representation of the experimentally used widefield microscope. Electron-beam lithography-fabricated Au-nanorods are probed by freely-diffusing IR895 in *n*-BuOH, which are illuminated in total internal reflection geometry and their emitted fluorescence is imaged onto an EMCCD camera. e) Simulated fluorescence time traces of molecules emitting in the vicinity of three different nanoantennas.

Here, we use the approach outlined above to extract molecule-antenna coupling information in a highly multiplexed fashion. We combine statistical data analysis of thousands of localization-events with systematic spectral-resonance scans of plasmonic nanorods to study the spectral-overlap dependent coupling of said structures to single molecules in the near-field. Furthermore, we show that rigourous statistical analysis of the localization error as well as the degree of linear polarisation (DOLP) gives direct access to the molecule-nanostructure interaction strength.

In order to image and study the interaction between single molecules and the said plasmonic nanostructures, we rely on a home-built total-internal-reflection (TIRF) microscope equipped with an NA 1.49 objective (Olympus APON60x) and an EM-CCD camera (Hamamatsu ImagEM X2) for widefield detection. We plane-wave illuminate the sample of interest with 10 kW/cm$^2$ of linearly-polarised 780 nm light under the critical angle. The objective lens collects the fluorescence emission which is separated from residual excitation light, by a dichroic beamsplitter in combination with an 808 nm longpass filter, and then imaged onto the EM-CCD camera at a magnification corresponding to 100 nm/px.

A typical sample consists of electron-beam lithography-fabricated Au nanorod-arrays with varying length and orientation (Figure 1c). Nanorod-molecule interactions and their effect on the molecules' apparent far-field locations are studied by adding a solution of the laser-dye IR895, contained in *n*-BuOH, on top of the array-sample and recording stochastically occurring fluorescence events. Here, TIRF illumination is crucially important as it allows selective interaction with the individual Au nanorods at the glass/*n*-BuOH interface while limiting the volume contributing to the fluorescence background to the extent of the evanescent field which is approximately 200 nm.

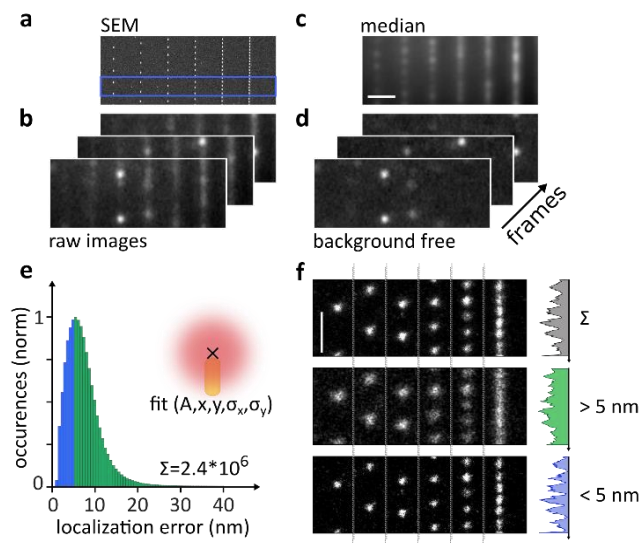

**Figure 2, Experimental procedure and proof of concept experiment.** a) SEM image of the studied array of 100x50x58 nm nanorods, with spacing decreasing down to 50 nm. b) Representative fluorescence images as recorded on the camera after adding the *n*-BuOH solution containing IR895. c) Median luminescence background computed from 400 images. d) Background-free fluorescence images obtained by subtracting the median image from the individual frames shown in (b). Scalebar (a)-(d): 1.5 µm. e) Localization error histogram at one standard deviation of all localizations registered in the proof-of-concept experiment. Inset: Schematic of 2D Gaussian fitting of individual emission events, yielding the relevant fitting parameters *A* (amplitude), *x* and *y* ( x- and y-position) and the respective localization errors σ$_x$ and σ$_y$. f) Antenna positions reconstructed from stochastic single-molecule localizations using: All localizations (top), only localizations with a localization error >5 nm (centre) and <5 nm (bottom). The area corresponds to the blue rectangle in a). Scalebar: 500 nm.

To gauge the overall experimental performance we fabricate 100x50x58 nm Au nanorods and decrease the apex-to-apex separation between individual rods down to 50 nm (Figure 2a), which is far below the diffraction limit at the detection wavelength >808 nm. The IR895 concentration is adjusted to approximately 10 µM leading to typical fluorescence images as shown in Figure 2b. Under the experimental conditions described above and an integration time of 50 ms we observe a constant fluorescence-background from rapidly-diffusing molecules, static photoluminescence from the individual nanorods reminiscent of the SEM-image convolved with the optical point-spread-function (PSF) and stochastically varying single-molecule emission events. The two static contributions are conveniently isolated by computing a temporal running median-image over batches of 400 images (Figure 2c). Subsequent subtraction of the median from the individual 400 video-frames yields essentially background-free images of the stochastically occurring emission-events of interest (Figure 2d).

Temporally, essentially all emission events last less than our integration time of 50 ms with only 2% exhibiting fluorescence intensity in consecutive frames. This observation suggests that no, or only very brief, binding of the molecule to the nanostructure takes place and that the overall "on-period" of individual molecules is considerably shorter than 50 ms. This is in good agreement with typical molecule transient times determined by fluorescent correlation spectroscopy measurements, which are on the order of tens to hundreds of microseconds, for a diffraction limited confocal volume, and even smaller for molecules diffusing through antenna hotspots[22,23].

At our IR895-concentrations typical widefield images exhibit only few, spatially well-separated, emission events per image and their nanometric position can be determined by Gaussian fitting (Figure 2e, inset). A typical localization-error histogram calculated as $\sqrt{\sigma_x^2 + \sigma_y^2}$ for 2.4*10$^6$ individual single-molecule localizations, obtained after a few-hours of continuous imaging at 20 frames-per-second, is shown in Figure 2e. Qualitatively, it is reminiscent of a Boltzmann distribution with a maximum around 5 nm. In the spirit of SRM we use the accumulated emission-events to reconstruct the underlying structure. We generate super-resolved images by summing area-normalised Gaussians with a width corresponding to their localization-error for all detection-events. This technique is often referred to as so-called *"PAINT"* microscopy in the biophysical community[24]. Figure 2f compares different reconstructions for the Au nanorod-array using all localization-events (top), localization events with large (centre) or small (bottom) localization error. A comparison of the SEM image and the reconstructed images shows good agreement between the real and the reconstructed nanorod distances (Figure 2a, highlighted area, and 2f) indicating that the fluorescence emission probability is greatly enhanced in the vicinity of the Au nanostructures. Indeed, using all detected events allows resolving individual nanorods separated by much less than the diffraction-limit. However, it is advantageous to exclusively rely on the subset of events with small localization errors in order to clearly resolve closely spaced nanorods.

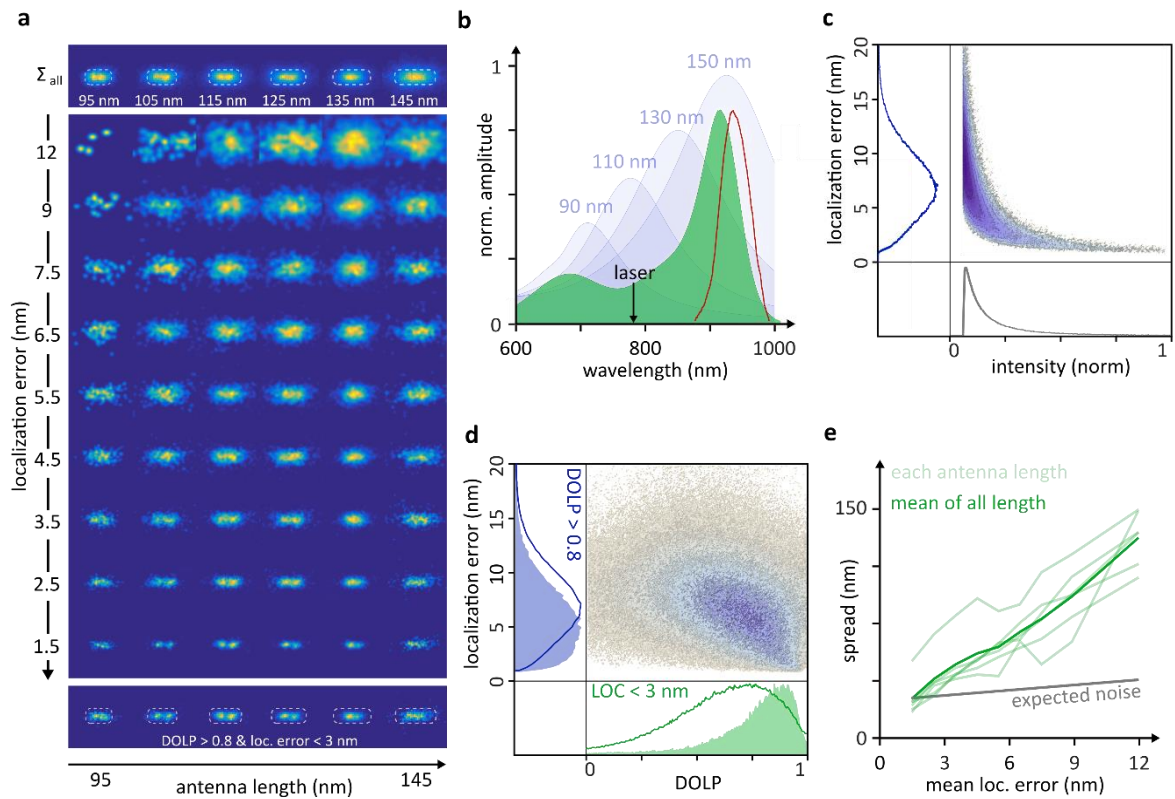

**Figure 3, Role of localization error and DOLP for mapping the nanoscale interaction.** a) Top: Antennas with lengths ranging from 95 to 145 nm reconstructed by relying on all detected fluorescence events. Centre: As before but by relying on data subsets with degree-of-linear-polarisation (DOLP)>0.8 and decreasing localization error and, Bottom: Final reconstruction with localization errors<3 nm and DOLP>0.8. 12 individual antennas are summed per antenna length. b) IR895 absorption (green) and fluorescence (red) spectra in comparison to simulated extinction spectra of Au nanorods of different length (blue). The excitation wavelength is indicated by the arrow. c) Correlation between emission intensity and localization error for all detected fluorescence events alongside histogram representations. d) Correlation between the DOLP and the localization error alongside histogram representations containing the full data-set (solid line) and the localization histogram for DOLP > 0.8 (blue, filled) and a DOLP histogram for a localization error < 3 nm (green, filled). e) Mean localization error dependence of the spread around the mean antenna position for all antenna lengths compared to the expected increase estimated based on shot-noise like localization deterioration.

After these proof-of-concept experiments we now move towards systematically mapping the interactions between Au nanorods and IR895 dye-molecules by tuning the nanorod resonances. More specifically, we fabricate nanorod-arrays with antenna length ranging from 95 nm to 145 nm and a width of 58 nm. We, furthermore, equip our detection channel with a Wollaston prism to allow polarisation-resolved detection parallel and perpendicular with respect to the long antenna-axis. Antenna-reconstructions obtained for s-polarised excitation and detection (along the long antenna axis), performed and analysed analogous to the initial proof-of-concept demonstrations, are shown in Figure 3a. These reconstructions are obtained by relying on all detected fluorescence events and, furthermore, averaging 12 individual antennas for each length to reduce the effect of possible nanofabrication-imperfections. Qualitatively, all interaction maps look identical albeit a gradual increase in size with increasing nanorod length. Importantly, all reconstructed areas are smaller than the nanorod itself, as expected for experiments performed in the vicinity of resonant metallic nanostructures which are known to displace the molecular emission in the far-field[14,16,17].

Based on the absorption and emission spectrum of the previously employed laser-dye IR895 (Figure 3b) with its strongly red-shifted fluorescence band centred around 920 nm, we would expect to observe distinct signatures of both excitation- as well as emission-enhancement phenomena due to the presence of the metallic nanostructures[15]. According to simulations (see Supplementary Material) the fluorescence emission maxima roughly overlaps with the resonance of the 145 nm nanorod, whereas the 780 nm excitation laser most efficiently interacts with the structures of approximately 110 nm in length. Consequently, we would expect well separated excitation- and emission-enhancement around 110 nm and 145 nm, respectively, which should manifest themselves in distinctly different antenna-reconstructions. However, none of the anticipated effects is observed in the interaction maps.

So far, we only used the localization information of the average of all events. To extract more specific field information, we perform statistical analysis of all localization events. Figure 3c shows a correlation between the fluorescence intensity and the localization error of all fitted emission events. We observe a shot-noise like correlation with an, approximate, square-root relation between localization error and fluorescence intensity. This observation indicates that neither strong background intensities nor possible PSF-distortions, due to strong molecule-antenna coupling, impact the experimental data as these effects would result in strong-departures from the anticipated behaviour. Therefore, these parameters contain qualitatively identical information and localization-event thresholding based on either parameter should, in principle, allow isolation of strongly interacting molecules. However, our previous observations suggest that the IR895 emission events are due to molecules merely diffusing through the antenna hotspots. As a result, localization error, or intensity, based thresholding might be unable to distinguish between transiently bound molecules in a region of medium intensity, or even on bare glass, and a molecule diffusing rapidly through the high-field-intensity region of interest.

To refine the extraction of the localization events of interest we therefore introduce a second parameter, the degree of linear polarisation (DOLP) computed as *DOLP=($I_s$-$I_p$)/($I_s$+$I_p$)* with $I_s$ and $I_p$ being the fluorescence intensity of individual molecules detected in the, simultaneously recorded, s- and p-polarised imaging channels, which are parallel and perpendicular with respect to the long antenna axis, respectively. We expect the DOLP to directly report on the antenna-molecule coupling and hence the molecules' position within the enhanced electric field around the nanorod. For molecules in close vicinity to a nanorod the emission should be strongly polarised along the antenna axis thus resulting in the largest possible DOLP. Figure 3d shows a DOLP-localization error correlation graph which indicates a qualitative correlation between the two values. As expected, we observe a shift towards larger DOLP values when filtering for events with small localization errors. We note that DOLP>0.8 can be readily obtained in our experiment as the molecules exhibit near-isotropic emission as they are freely diffusing in *n*-BuOH with a refractive index of $n_{nBuOH}$=1.4 which is close to the n=1.52 of the substrate. As a result, the commonly observed depolarisation for high-NA objectives is far less pronounced[25,26].

Based on the correlation-graphs, we now attempt a refined antenna-reconstruction. We keep the DOLP>0.8 to select molecules that show considerable antenna-coupling and evaluate different localization error intervals. For the most precise localization we observe a pronounced double-spot pattern with an approximate spot-to-spot separation of 35 nm, for shorter antennas, which merges into a single, slightly elongated spot, for longer antennas (Figure 3a, centre). As we move towards higher localization errors the features disappear, even for errors as small as 4.5 nm which is surprising given the spot-separation of 35 nm. Intuitively, we would assume that any super-resolved feature is blurred by the localization error but a mere increase of 4.5-2.5=2.0 nm results in complete disappearance of the, previously, clearly resolved double-spots. To rationalise this surprising observation we quantify the deterioration by computing the spread around the mean antenna position $(\bar{x}, \bar{y})$ as:

$$\frac{\sum_{i=0}^{N} \sqrt{(x_i - \bar{x})^2 + (y_i - \bar{y})^2}}{N}$$

Here, we observe a dramatic departure from the expected shot-noise-limited like increase (Figure 3e) with the spread around the antenna mean increasing by 124 nm for a localization error increase of 10.5 nm. From a SRM point-of-view this observation might appear surprising. However, the electric field distribution around a nanoantenna is not a structure with a well-defined, binary, shape but rather a function with boundaries that depend on the intensity-enhancement and interaction strength with the molecular probes. A poorly-localized molecule is therefore more likely to be further away from the nanorod than a well-localized one thus increasing the absolute area that is sampled around the antenna. This localization error based sampling space results in a rapid loss of the high electric-field-intensity information. Based on these observations we attempt the "ideal" antenna intensity reconstruction based on all localization events with a DOLP>0.8 and a localization error <3 nm (Figure 3a (bottom)), which shows a gradual transition from a double to a single spot upon varying the antenna resonance.

To rationalise these experimental observations we perform finite-difference-time-domain (FDTD) simulations using the Lumerical solver (see Supplementary Material). To this end we simulate the radiation patterns of x- and y-oriented dipoles (molecules) at different positions around the nanorod and then generate far field images reflecting our experimental imaging system. The respective images are processed analogous to the experimental data. Ultimately, we obtain a simulated reconstructed image where the intensity of each dipole emitter corresponds to the excitation field intensity of the nanorod at the emitters' location. Figure 4a shows a comparison between the experimentally obtained results for 115 nm and 135 nm long nanorods and the simulation. Here, the two length were chosen as being representative candidates for excitation- and emission-enhancement, respectively. As in the experiment, we observe a the pattern change with most of the fluorescence intensity being localized in the centre of the antenna once its resonance overlaps with the molecular emission spectrum which is in good agreement with previously reported experimental studies and simulations[12,27]. The slight deviations between our experimental data and the simulations are most likely due to the insufficient quality of the experimental data as well as the crude modelling of the antenna-molecule coupling strength. Here, we restrict our model two a single 2D plane, located 5 nm above the sample-substrate, of molecule-nanorod interactions rather than the full 3D volume around the antenna. Additionally, we model the ideal shape of the nanorod after fabrication, where surface irregularities and slight deformations are ignored, an ideal scenario that we tried to reproduce by summing a total of 12 individual antennas per length. Especially for excitation wavelengths blue-shifted with respect to an antenna resonance the excitation enhancement is generally reduced[28,29] thus complicating the experiment considerably, a fact that manifests itself in the reduced data quality of the 135 nm data as compared to the data obtained with the red-shifted excitation of the 115 nm antenna.

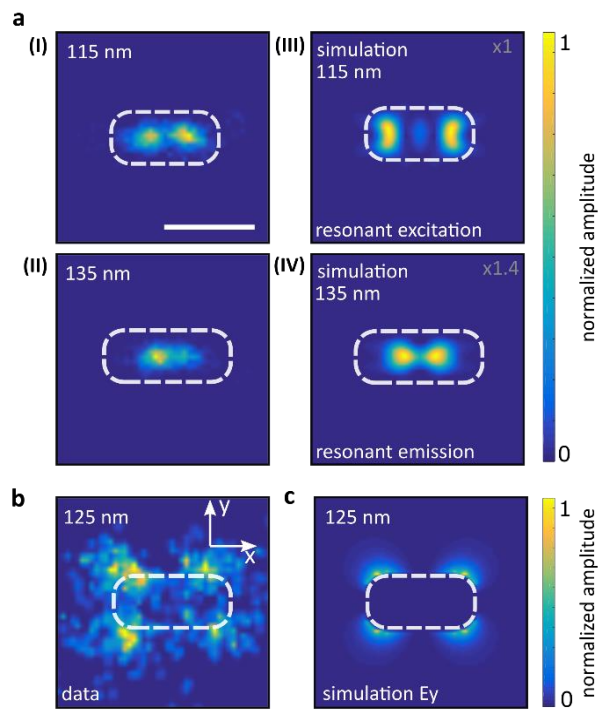

**Figure 4, Comparison between experimental observations and simulations for all polarization configurations.** a) Experimentally reconstructed intensity distribution around 115 nm and 135 nm antennas by applying a DOLP>0.8 and localization error<3 nm threshold (left) in comparison to the distribution expected from FDTD simulations (right). b) Experimentally reconstructed intensity distribution for cross-polarised localizations with localization errors<7.1 nm and DOLP<(-0.4) in comparison to an FDTD simulation of the y-component of the electric field around the nanoantenna.

To conclude the data-driven mapping of molecule-antenna coupling strength we now turn our attention to the molecular events detected with emission polarization perpendicular with respect to the excitation polarisation as well as the antenna-resonance axis. Based on the strong DOLP bias of our experiment (Figure 3d) and the unfavourable excitation and emission enhancement of perpendicularly-oriented molecules we expect few emission events with little-to-no intensity in the cross-polarised imaging channel. The large amount of localization events recorded nevertheless allows us to analyse the electric fields governing emission through this channel. Figure 4b shows an emission-intensity reconstruction of the 125 nm nanorod for 12 summed antennas based on approximately 1100 events with a localization

error<7.1 nm and DOLP<(-0.4). The pattern is reminiscent of the cloverleaf-like PSFs previously observed as far-field emission patterns of individual emitters positioned above a nanowire[30]. However, our reconstruction is based on post-detection reconstruction which incorporates Gaussian fitting which necessarily results in a complete loss of PSF shape-information. To rationalise our observation we simulate the y-component of the electric field present around a nanorod after far-field TIRF-excitation with light polarised along the long axis of the nanorod. The near-perfect agreement between experimentally observed fluorescence localizations and theoretically predicted excitation intensity suggests that we purely probe the y-component of the electric antenna field without any interaction of molecule and nanoantenna. Our results confirm the findings obtained by direct near field imaging of the vectorial nanorod fields with a scanning antenna tip[18].

In summary, we constructed a widefield TIRF microscope and employed arrays of nanoantennas with increasing length that resonantly interact with individual emitters at both excitation and emission frequencies. Depending on the antenna length either emission- or excitation-induced effects resulted in changes in the observed super-resolved intensity reconstructions as previously observed in a variety of studies. Instead of concentrating on these now well understood, and widely studied, discrepancies between expected and measured electric field distributions we, instead, emphasise the true strength of widefield super-resolution mapping of plasmonic structures: the possibility to perform highly-parallelised high-throughput experiments in combination with thorough statistical analysis. While most of nanoplasmonics relies on point-scanning, or confocal, observations which are intrinsically throughput-limited our method yields millions of single-molecule observations while, simultaneously, being considerably easier to experimentally implement. As a result, intensity and polarisation measurements allow us to isolate a small subset of strongly-coupled single molecules from an enormous amount of less-coupled emitters and background, purely based on statistical analysis that relies on physically meaningful expectations. Based on these analysis we resolved the expected double-spot excitation enhancement pattern present around the antenna-apexes which would have remained unresolved if statistical filtering would not have been applied. Furthermore, we were able to resolve the $E_y$-component of the electric field in a cross-polarisation analysis, by isolating the few events that did not couple to the long axis of the nanoantenna.

Our approach highlights that single-molecule based approaches are extremely powerful when conducted at ensemble-level detection levels. Here, rarely occurring events, such as the strong coupling of an individual molecule to a plasmonic structure, can be identified with statistical significance. Importantly, such observations are impossible with ensemble-based techniques and would have most likely been discarded as artefacts in traditional single molecule experiments which often draw conclusions based on a few tens of observations. Ensemble-level single molecule spectroscopy eliminates these problems and allows confident identification and study of physical phenomena based on rigorous statistical analysis.

We are currently extending our methodology towards highly-multiplexed imaging incorporating simultaneous position, polarisation and spectral detection. These extended capabilities, in combination with the statistical analysis introduced here, will allow us to observe the transition from uncoupled to strongly coupled emitters as they approach resonant plasmonic nanostructures.

**Supplementary Information**

*FDTD calculations (I) – Molecule and antenna interaction.* The FDTD calculations are performed using the Lumerical FDTD solver. In all calculations, the nanostructure is modelled as being supported on top of a glass substrate with a refractive index of 1.5 and being surrounded by a medium with a refractive index of 1.4, corresponding to the refractive index of *n*-BuOH. The nanostructure is a nanorod with a constant width of 58 nm and height of 50 nm, and a length of 115 nm or 135 nm, on top of a 2 nm thick titanium layer. We simulate the radiation patters of X and Y oriented dipoles at a fixed height of 5 nm above the glass substrate at different positions around the nanorod and collect the electric fields 10 nm below the substrate. We then perform a Fourier transformation to access the far field. Only collection angles smaller than the critical angle defined by our numerical aperture of 1.49 are retained. After Fourier filtering the radiation pattern is propagated back into the image plane and fitted analogous to the experimentally obtained data. A 2D Gaussian is then plotted at each obtained x-/y-position using a width of 6 nm at one standard deviation and an intensity corresponding to the excitation intensity of the nanorod at the original location of the dipole emitter – here we match the two dipole orientations with the two components of the excitation field ($E_x$ and $E_y$).

*FDTD calculations (II) - Nanorod resonances.* To simulate the nanorod resonances in TIRF excitation and submerged in butanol, we used the total-field-scattered-field source in order to quantify absorption and transmission of the plasmonic structure at different wavelengths.

*Sample preparation.* We use electron-beam lithography to fabricate the sample. To this end, we evaporate a 2 nm titanium layer followed by a 50 nm gold layer. We spincoat 150 μm of the resist ARN7500.08 on top of the gold at 8000 RPM for 60 s. After electron-beam exposure and development (ARN300, diluted 4:1 with water), reactive Ion etching is employed with an Oxford Instruments Plasmalab System leaving only the Au nanorods on top of the coverglass.